\documentclass[useAMS,usenatbib]{mn2e}
\usepackage{graphicx}
\usepackage{amsmath}
\usepackage{color}

\newcommand{\Teff}{\mbox{$T_{\mathrm{eff}}$}}

\newcommand{\Msun}{\mbox{$\mathrm{M}_{\odot}$}}

\newcommand{\ugr}{\mbox{($u-g)$ vs. $(g-r)$}}

\title{The incidence of magnetic fields in cool DZ white dwarfs}

\author[M.A. Hollands]{
M.A. Hollands$^1$,
B.T. G\"ansicke$^1$,
D. Koester$^2$\\
$^{1}$ Department of Physics, University of Warwick, Coventry CV4 7AL,
UK \\
$^{2}$ Institut f\"ur Theoretische Physik und Astrophysik, University of Kiel,
24098 Kiel, Germany\\
}

\begin{document}

\date{Accepted 2015 March 12.}

\pagerange{\pageref{firstpage}--\pageref{lastpage}} \pubyear{2015}

\maketitle

\label{firstpage}

\begin{abstract}
Little is known about the incidence of magnetic fields among the coolest white dwarfs.
Their spectra usually do not exhibit any absorption lines as the bound-bound opacities of hydrogen and helium are vanishingly small.
Probing these stars for the presence of magnetic fields is therefore extremely challenging.
However, external pollution of a cool white dwarf by, e.g., planetary debris, 
leads to the appearance of metal lines in its spectral energy distribution.
These lines provide a unique tool to identify and measure magnetism in the coolest and oldest white dwarfs in the Galaxy.

We report the identification of 7 strongly metal polluted,
cool ($T_{\mathrm{eff}}<8000$\,K) white dwarfs with magnetic field strengths ranging from 1.9 to 9.6\,MG.
An analysis of our larger magnitude-limited sample of cool DZ yields a lower limit on the magnetic incidence of
$13\pm 4$\,percent, noticeably much higher than among hot DA white dwarfs.

\end{abstract}


\section{Introduction}
White dwarfs (WDs) have been known to harbour magnetic fields since the detection of circularly polarised light
from GJ~742 \citep{kempetal70-1}.
In the following decades a plethora of magnetic WDs (MWDs) have been identified either from Zeeman splitting
of absorption lines in their spectra or by spectropolarimetry \citep[and references therein]{kawkaetal07-1}.
A wide variety is seen in temperature, atmospheric composition, and field strength.
The advent of large scale spectroscopic surveys, in particular the Sloan Digital Sky Survey (SDSS),
has in the last decade increased the number of known MWDs to several hundred
\citep{gaensickeetal02-5,schmidtetal03-1,vanlandinghametal05-2,kleinmanetal13-1,kepleretal13-1,kepleretal15-1}.

Despite the ever growing list of these previously rare objects,
two questions continue to remain without a definite answer:
What is the origin of these magnetic fields?
And what is the fraction of WDs that are magnetic, and how does this vary with cooling age/temperature?

Two distinct models have been proposed to explain the emergence of fields $\ga1$\,MG
in isolated WDs.
In the fossil field hypothesis, the magnetic fields of the chemically peculiar Ap/Bp stars are thought
to be amplified due to flux conservation during post-main sequence evolution resulting in
WDs with fields in the MG regime \citep{woltjer64-1,angeletal70-1,angel81-1,wickramasinghe+ferrario00-1}.
A more recent hypothesis \citep{toutetal08-1} considers a binary origin,
where a system undergoing a common envelope leads to magnetic dynamo generation.

The incidence of magnetism in WDs remains poorly estimated due to selection effects.
Independent studies are difficult to reconcile with one another as each suffers from its own set of biases.
This problem becomes significantly more pronounced when focusing on subsets of the total WD population
where small number statistics dominate.
Recent volume limited samples of nearby WDs present the most unbiased estimates of the magnetic
incidence when considering all WD sub types,
and suggest incidences of $21\pm8$\,percent for WDs within 13\,pc of the Sun,
and $13\pm4$\,percent for those within 20\,pc \citep{kawkaetal07-1}.
However these MWDs are dominated by fields lower than $100$\,kG and strongly magnetic objects with fields above $10$\,MG.
Only $1$ out of the 15 MWDs in the compilation of \citet{kawkaetal07-1} has a field strength between 1 and 10\,MG
(the range that we discuss in this work).
More recently, \citet{sionetal14-1} have presented a volume limited WD sample within 25\,pc from the Sun.
They find a magnetic incidence of $8$\,percent when considering magnetic fields above 2\,MG only.
Other studies have investigated the magnetic incidence with much larger, but magnitude-limited samples.
For instance \citet{kleinmanetal13-1} identified over 12000 DAs\footnote{WDs showing only hydrogen/helium
lines in their spectra are classified DA/DB, with only metal lines as DZ, and without any spectral lines as DC.
Magnetic DA WDs where magnetism is detected via Zeeman splitting are known as DAH, and via polarimetry
as DAP \citep{sionetal83-1}.}
from SDSS data release 7 (DR7) spectra,
of which over 500 are suggested to be magnetic \citep{kepleretal13-1}, leading to a much lower incidence of 4\,percent.
However, because this sample is magnitude-limited, it is intrinsically biased.
Most degenerates in the local sample have temperatures below 10000\,K, whereas 84\,percent of the WDs from \citet{kepleretal13-1}
are hotter than this.
The discrepant numbers between the local sample of cool/old WDs and hotter/younger WDs,
have been the basis for some authors to claim an age-dependency of the magnetic incidence \citep{fabrikaetal99-1,liebertetal03-1}.

Analysing the small sample of WDs with accurate parallaxes,
\citet{liebertetal88-1} noted that magnetic WDs appear to be under-luminous for their colour,
suggesting they have smaller radii, and hence higher masses, than non-magnetic WDs.
Later, \citet{liebertetal03-1} derived a mean mass of 0.93\,\Msun\ for eight MWDs from the
Palomar Green (PG) survey, based on model atmosphere analyses, compared to ${\sim}0.6$\,\Msun\ for non-magnetic ones.
While there is hence independent evidence for higher-than-average masses for MWDs,
caveats to bear in mind are that there are still few MWDs with precise parallaxes,
and even for those systematic uncertainties in the analysis of their spectra limits the accuracy
of the desired masses \citep{kulebietal10-1}.

A common theme among all the above investigations is that the true magnetic incidence is expected be higher,
as the various biases (e.g. signal-to-noise, magnetic broadening) tend to work against the identification of MWDs.

The very coolest WDs (\Teff$<8000$\,K) do not show optical lines of hydrogen or helium
since these elements are in their ground states in the low temperature atmospheres.
This transition to featureless (DC) spectra occurs at around 11000\,K for WDs with helium dominated atmospheres,
and 6000\,K for WDs with hydrogen dominated atmospheres \citep{vauclairetal81-1,bergeronetal01-1}.
Because of this absence of absorption lines, it is not possible to identify magnetism in these stars
via Zeeman splitting.
A handful of DCs have been found to be magnetic through spectropolarimetry \citep{putney97-1}.
However, this method is expensive and is unsuitable for most known cool WDs because of their faintness.

\section{Metal polluted white dwarfs}
\label{metalWD}The last two decades have seen great interest devoted to the study of WDs with atmospheres contaminated by metals.
The now widely accepted scenario is that after post-main sequence evolution to the WD stage,
an accompanying planetary system will undergo dynamical instability \citep{debesetal02-1,verasetal13-1}.
Small rocky bodies such as asteroids or minor-planets may then have their orbits perturbed by a larger
planetary object and soon find themselves venturing into the Roche-radius of the WD,
resulting in their tidal disruption \citep{debesetal12-1,verasetal14-1} and subsequent formation of a circumstellar debris disk.
This material is then accreted onto the star \citep{jura03-1,rafikov11-1} producing metal lines 
in the spectra of these WDs.

Cool WDs displaying only metal lines are classified as DZ.
\citet{farihietal10-2} showed that DZ and DC white dwarfs share the same velocity, spatial and temperature distributions
and should therefore belong to the same stellar population.
Assuming the existence of magnetism among DZ stars is uncorrelated with the presence of metals,
the detection of split metal lines in the spectra of DZ white dwarfs becomes a powerful tool for determining
the magnetic field strength distribtion among the very coolest and oldest of WDs.
However, prior to this work, only 3 magnetic DZ WDs were known:
LHS2534 \citep{reid01-1}, WD0155+003 \citep{schmidtetal03-1}, and G165-7 \citep{dufouretal06-1},
with respective surface averaged field strengths, $B_S$, of $1.9$, $3.5$, and $0.6$\,MG.
These all have SDSS spectra and so we refer to these using the SDSShhmm$\pm$ddmm naming format
(SDSS1214$-$0234, SDSS0157$+$0033, and SDSS1330$+$3029 respectively) for consistency with our new
identifications and the names used in \citet{koesteretal11-1}.

Recently, \citet{koesteretal11-1} identified a sample of 26 cool (\Teff$<9000$\,K) DZ with strong photospheric
metal pollution, filling previously empty parameter space at low \Teff\ and high atmospheric Ca abundance
compared with the preceding work by \citet{dufouretal07-2}.
The rocky nature of the accreted material is evident from the variety of detected metals which includes
Ca, Mg, Fe, Na, Cr, and Ti, and the low abundance of H.
The \citet{koesteretal11-1} sample also includes re-identifications of the known magnetics SDSS0157$+$0033
and SDSS1330$+$3029 (see above).

\section{results}

\subsection{Identification}
\label{ident}
\begin{figure}
  \centering
  \includegraphics[angle=0,width=\columnwidth]{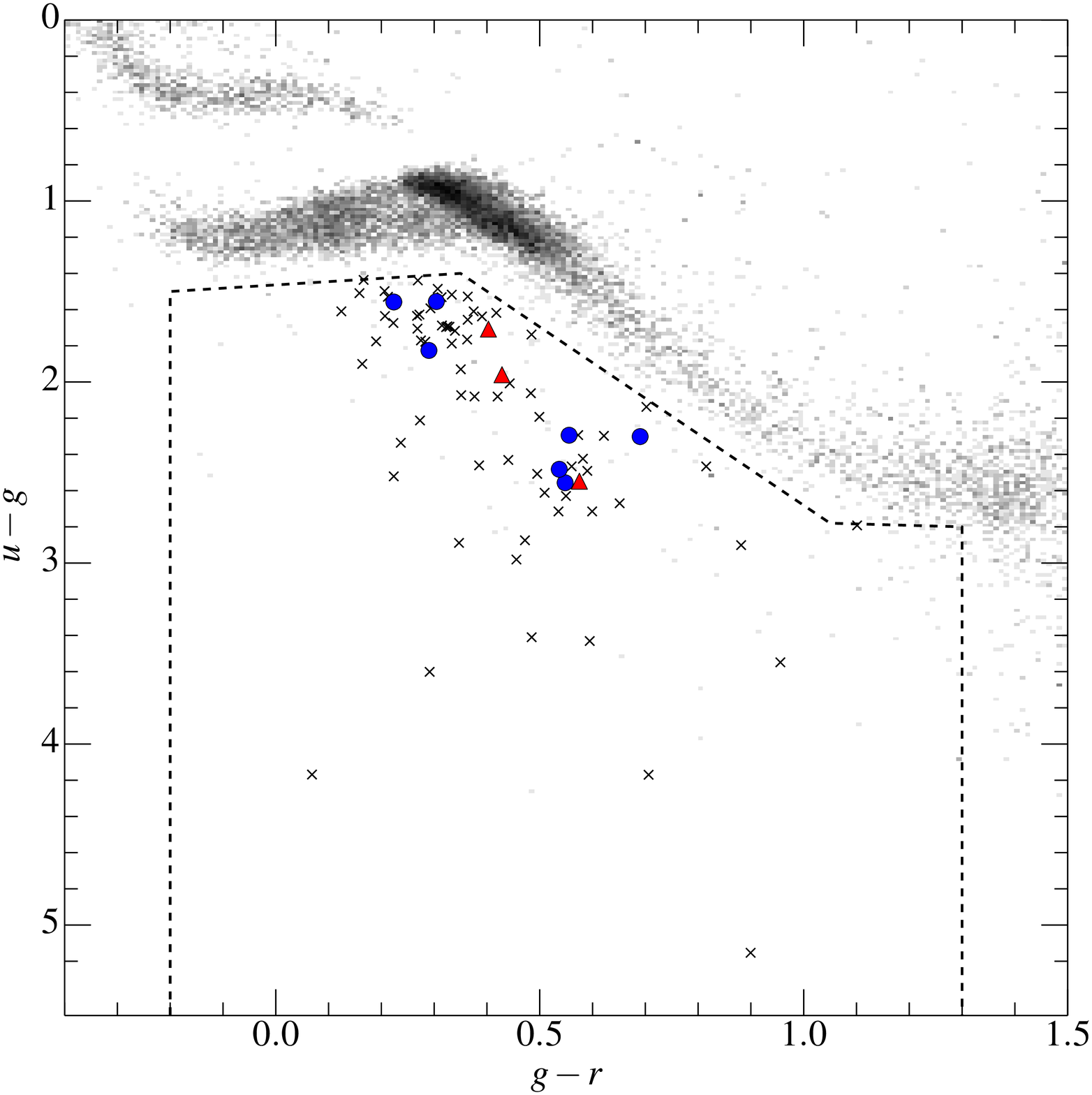}
  \caption{\label{mag_ccd}
   The colour-cut used to identify cool DZ WDs extends below the main sequence (dashed line).
   Previously known magnetic DZ are indicated by filled triangles,
   and new discoveries by filled circles.
   Crosses indicate the other 69 DZ in our sample of 79 which do not show Zeeman splitting
   of spectral lines.
   The main sequence is indicated by the grey-scale.}
\end{figure}

\begin{figure*}
  \centering
  \includegraphics[angle=0,width=0.99\textwidth]{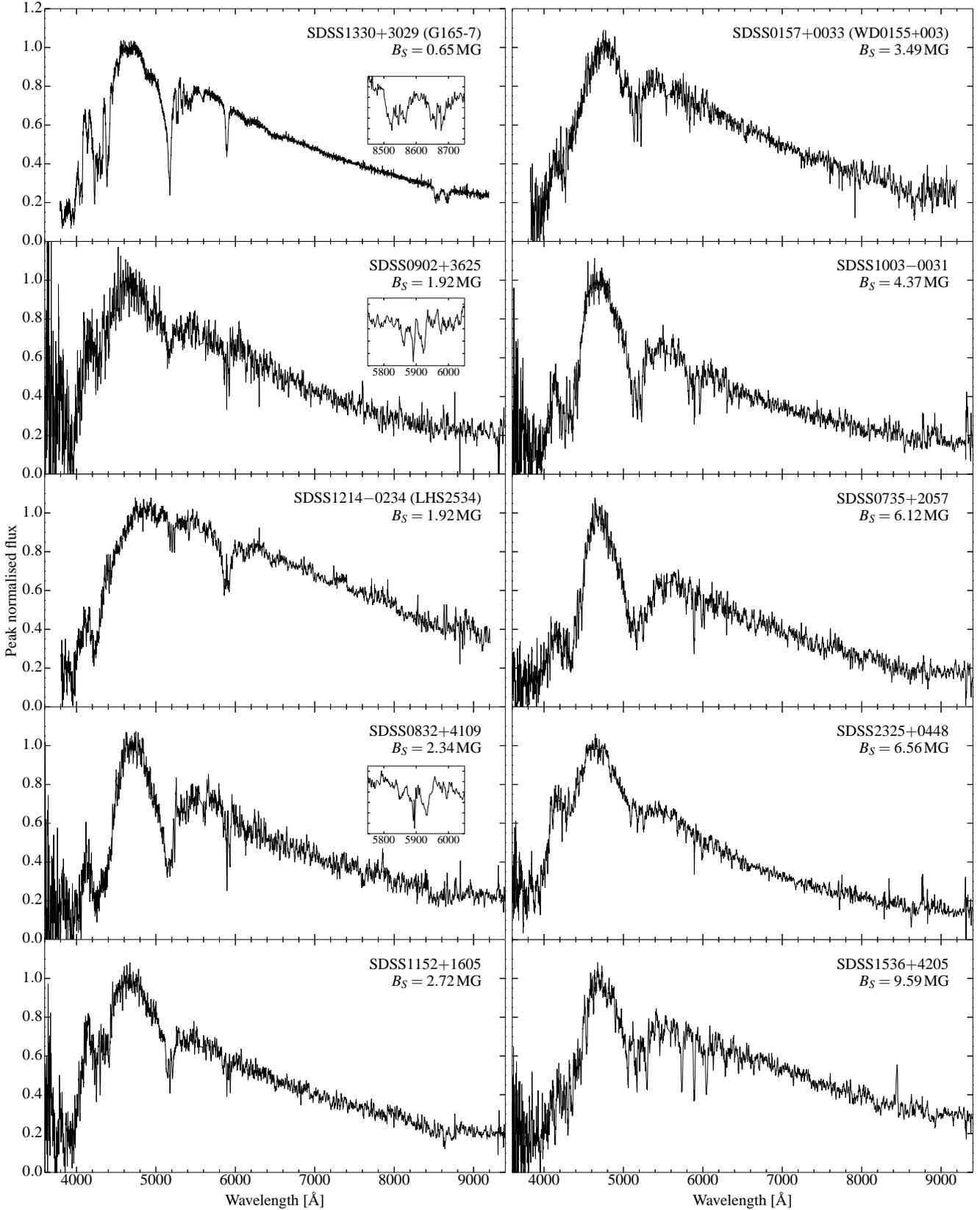}
  \caption{\label{all_spec}
  Optical SDSS spectra for all 10 magnetic DZ, ordered from top to bottom by increasing magnetic field strength.
  Data are smoothed by a 7-point boxcar to reduce the noise level and enhance spectral features,
  with the exception of the bright object SDSS1330$+$3029.
  Insets are plotted where splitting is not obvious in the full spectrum (same smoothing applied).
  }
\end{figure*}

In an extension of the work by \citet{koesteretal11-1} using the more recent DR10 spectroscopic data,
we have identified a sample of 79 strongly polluted cool DZ, with the serendipitous discovery that
10 are magnetic.
Full details of this sample will be presented elsewhere (Hollands et al. in prep);
here we only provide a brief summary of the selection procedure.

Firstly, we queried SDSS Casjobs \citep{lietal08-1} for objects with colours below the main sequence in the \ugr\
two-colour diagram (Figure~\ref{mag_ccd}).
This region of colour-space is highly abundant in quasars of redshift 2--5.
To remove these, we employed two independent cuts in proper-motion and redshift:
a minimum proper-motion threshold of 3-$\sigma$ above zero and
SDSS spectroscopic redshift below $0.01$.
Any spectrum with non-zero redshift \textsc{zwarning} (0 indicates no warning) was also kept.

Requiring the objects to pass at least one of the proper-motion or redshift selection criteria was necessary
as neither could be completely relied upon in isolation.
On the one hand, proper motions may be unavailable if the object is too faint to be detected in the USNO-B images
used to detect positional variations.
Additionally, because detectable cool DZ are by definition relatively nearby (as they are intrinsically faint),
some will have sufficiently high proper-motions to evade a match between SDSS and USNO-B observations.
On the other hand, the unusual spectra of DZ WDs with their deep, broad metal lines can trick
the SDSS redshift pipeline into classifying them as quasars, reporting values of $z>1$, with \textsc{zwarning} $=0$.

In addition to quasars, low mass main sequence stars strayed into our colour cut and were removed via
template fitting for K and M spectral types.
Finally, the DZ WDs were identified by visual inspection of the remaining spectra.
This method recovered all 17 DZ WDs presented by \citet{koesteretal11-1} that fell into our colour-cut,
while also finding 62 additional objects, bringing our total colour-selected sample to 79.
Magnetic DZ were identified from splitting of the Mg~\textsc{i} blend at 5172\,\AA\ and the
Na\,\textsc{i} doublet at 5893\,\AA.

In total we found our sample of 79 DZ contains 10 MWDs,
which are listed in Table~\ref{tab:mag_DZ} with their SDSS optical spectra shown in Figure~\ref{all_spec}.
Among these 10 objects, 3 are the previously known magnetic DZ as mentioned in \S\ref{metalWD}
(SDSS0157$+$0033, SDSS1214$-$0234, and SDSS1330$+$3029).
One system, SDSS1152$+$1605, was first identified as a DZ by \citet{koesteretal11-1},
who also pointed out the possible magnetic nature of the star.
However the signal-to-noise (S/N) of the single SDSS spectrum available was too low for a firm conclusion.
The newer SDSS spectrum, presented here, shows unambiguous Zeeman splitting.
The remaining 6 objects are all newly identified WDs.
We note that the recently published WD list of \citet{kepleretal15-1} independently identifies
SDSS0735$+$2057, SDSS0832$+$4109, SDSS1003$-$0031, and SDSS1536$+$4205,
however only SDSS1536$+$4205 is reported to be magnetic.

\begin{table*}
  \centering
  \caption{\label{tab:mag_DZ} Magnetic DZ with field strengths, temperatures and SDSS PSF photometry.
  Previously known magnetic DZ are indicated by their starred coordinates.
  The field strengths for SDSS1214$-$0234 and SDSS1330$+$3029 are taken from \citet{reid01-1} and \citet{dufouretal06-1} respectively.
  The quoted $B_S$ uncertainties are obtained from the formal errors on our fit parameters. Where a good fit is obtained for both
  Mg and Na lines, their weighted mean is used.}
  \begin{tabular}{lccccccc}
    \hline
    \hline
    SDSS & $B_S$ (MG) & \Teff\ (K) & $u$ (mag) & $g$ (mag) & $r$ (mag) & $i$ (mag) & $z$ (mag) \\
    \hline
    J133059.26$+$302953.2* & $0.65$          & 6000 & $18.28\pm 0.02$ & $16.32\pm 0.02$ & $15.89\pm 0.01$ & $15.91\pm 0.02$ & $16.09\pm 0.03$\\
    J090222.98$+$362539.6  & $1.92\pm 0.05$  & 6300 & $22.58\pm 0.32$ & $20.75\pm 0.03$ & $20.46\pm 0.04$ & $20.36\pm 0.06$ & $20.50\pm 0.22$\\
    J121456.39$-$023402.7* & $1.92$          & 5200 & $20.87\pm 0.06$ & $18.32\pm 0.02$ & $17.75\pm 0.01$ & $17.56\pm 0.01$ & $17.51\pm 0.02$\\
    J083200.38$+$410937.9  & $2.35\pm 0.11$  & 5900 & $23.62\pm 0.79$ & $21.32\pm 0.05$ & $20.63\pm 0.03$ & $20.70\pm 0.05$ & $20.75\pm 0.14$\\
    J115224.51$+$160546.7  & $2.72\pm 0.04$  & 6500 & $21.73\pm 0.12$ & $20.18\pm 0.03$ & $19.95\pm 0.02$ & $20.02\pm 0.03$ & $20.08\pm 0.09$\\
    J015748.14$+$003315.0* & $3.49\pm 0.05$  & 5700 & $21.30\pm 0.07$ & $19.59\pm 0.02$ & $19.19\pm 0.02$ & $19.21\pm 0.02$ & $19.36\pm 0.05$\\
    J100346.66$-$003123.1  & $4.37\pm 0.05$  & 6300 & $22.91\pm 0.33$ & $20.61\pm 0.03$ & $20.06\pm 0.02$ & $20.00\pm 0.03$ & $20.20\pm 0.11$\\
    J073549.19$+$205720.9  & $6.12\pm 0.06$  & 6000 & $23.08\pm 0.34$ & $20.53\pm 0.02$ & $19.98\pm 0.02$ & $19.92\pm 0.02$ & $20.09\pm 0.09$\\
    J232538.93$+$044813.1  & $6.56\pm 0.09$  & 7200 & $21.44\pm 0.15$ & $19.88\pm 0.02$ & $19.58\pm 0.03$ & $19.68\pm 0.04$ & $19.69\pm 0.11$\\
    J153642.53$+$420519.2  & $9.59\pm 0.04$  & 5500 & $23.32\pm 0.65$ & $20.84\pm 0.04$ & $20.30\pm 0.03$ & $20.17\pm 0.04$ & $20.33\pm 0.16$\\
    \hline
  \end{tabular}
\end{table*}

\subsection{Average magnetic field measurement and effective temperatures}
\label{average}
In the spectra of all 10 objects (with the exception of SDSS1330$+$3029 where Zeeman splitting is only seen in the Ca triplet),
we are able to identify split lines of either Mg~\textsc{i}, Na~\textsc{i}, or both.
For field strengths of $\ga2$\,MG,
spin and orbit angular momenta decouple, and so the Paschen-Back approximation is appropriate.
Therefore we treat the splittings as triplets.

We used a 7-parameter fit to the observed line profiles to measure the field strength.
The continuum flux in the vicinity of the triplet was modelled as a quadratic in $F_\nu$.
A linear approximation would not suffice, particularly for the wings of the broad Mg feature (see Figure \ref{all_spec}).
We then modelled the triplet as the sum of three Gaussian profiles with equal width,
depth (in continuum normalised flux) and separation in wavenumber, $1/\lambda$.
The wavenumber of the $\pi$-component of the triplet was also included as a free parameter to account for small shifts.
In all cases, we found the $\pi$-components are blue shifted from their rest wavelengths,
and generally increasing with field strength, suggesting this is predominantly caused by the quadratic Zeeman effect,
with only minor contributions from gravitational and Doppler shifts.
The maximum blueshift of the $\pi$-component is found to be 5\,\AA\ for the Na triplet of SDSS1536$+$4205.
The small (few percent) measurement error this may have on our field measurements does not affect
our discussion on magnetic incidence.

We used least-squares minimisation via the Levenberg-Marquardt algorithm to optimise these parameters.
Where possible we fitted both the Mg and Na lines, however, this could not always be achieved for a variety of reasons:
One of the lines may be significantly less deep than the other;
the Mg line  in some cases is very broad and asymmetric, such that the 3 components cannot be distinguished;
or poor subtraction of sky emission distorts the flux near the $\pi$ component
of the Na triplet, making a fit to this line less reliable than for the Mg triplet.

\begin{figure}
  \centering
  \includegraphics[angle=90,width=\columnwidth]{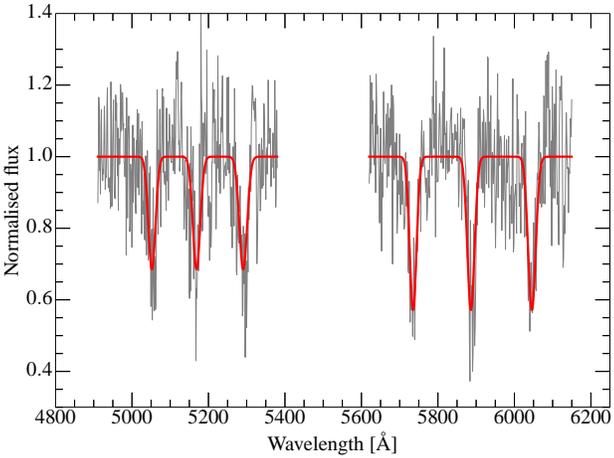}
  \caption{\label{1536_fit}
  Fits to the Mg and Na splittings for SDSS1536$+$4205.
  Lines are fit with Gaussians with equal $1/\lambda$ separations from the central $\pi$ components.}
\end{figure}

The average surface magnetic field strength, $B_S$, was subsequently calculated from
\begin{equation}
B_S/\mathrm{MG} = \frac{\Delta(1/\lambda)}{46.686},
\label{eq:Bsep}
\end{equation}
where $\Delta(1/\lambda)$ is the inverse wavelength separation in cm$^{-1}$ between the components of a triplet \citep{reid01-1}.
As an example, the fit to the most magnetic object SDSS1536$+$4205 is shown in Figure~\ref{1536_fit} with a measured field
strength of $B_S=9.59\pm0.04$\,MG.

Additionally, we estimated \Teff, by fitting each SDSS spectrum to a grid of DZ
model spectra\footnote{As our non-magetic models ignore the potential additional line blanketing from the
Zeeman-split lines, the true effective temperatures may differ from the values quoted here}.
The grid was calculated for temperatures from $4400$ to $14000$\,K in steps of $200$\,K, and Ca
abundances from $\log_{10}$[Ca/He] = $-10.50$ to $-7.00$ in steps of $0.25$\,dex.
Because photospheric metal lines arise from the accretion of planetary debris
\citep{zuckermanetal07-1,gaensickeetal12-1},
all other metals were held at a fixed ratio to Ca, assuming bulk Earth abundances.
The models were calculated for fixed surface gravity of $\log g = 8.0$.

The SDSS spectra were fit to each of the model spectra to obtain a $\chi^2$, at each grid point.
The $\chi^2($\Teff$,\log_{10}[\mathrm{Ca}/\mathrm{He}])$ plane was then interpolated with
bicubic splines to more accurately locate the best fit in the parameter space.
The corresponding effective temperatures are quoted in Table~\ref{tab:mag_DZ} to the nearest 100\,K.

\subsection{Field measurement for an offset dipole}
\label{offset_dipole}
For a MWD with the simplest possible field structure, a centred magnetic dipole,
the magnetic field across the star varies with magnetic latitude,
resulting in a field twice as strong at the poles compared to the magnetic equator \citep{achilleosetal92-2}.
A spectrum taken of a MWD integrates over its entire visible surface, and therefore over the range in field strengths,
resulting in  magnetic broadening of the $\sigma$ components in a given Zeeman triplet.
This effect is generally observed in  magnetic WDs and can be used to constrain the inclination
to the magnetic axis to some degree \citep[e.g.][]{bergeronetal92-3}.

A brief inspection of Figure~\ref{all_spec} shows that some of the MWDs in our sample have
very narrow Zeeman triplets, where all three components have similar depths.
This is seen in SDSS0157$+$0033, SDSS1003$-$0031, SDSS2325$+$0448, and in particular SDSS1536$+$4205.
A centred dipole field observed from any inclination to the magnetic axis can not reproduce these line profiles.

This suggests that these WDs may have more complex field topologies.
We show that a dipole offset from the star's centre can reproduce the observed Zeeman line profiles.
In principle this offset, $a$, can be in any direction relative to the unshifted magnetic field axis
\citep{achilleosetal89-1},
however, here we consider only displacement along the magnetic dipole axis (which we define to be in the $z$-direction),
i.e. $a_x=a_y=0$ as in \citet{achilleosetal92-2}.

For an arbitrary point on the surface of the WD with coordinates $(x,y,z)$ in units of $R_\mathrm{WD}$,
the strength of the field, $B(x,y,z)$, is given by \citep{achilleosetal92-2}
\begin{equation}
  B(x,y,z) = B_d\left[ r^2 + 3(z-a_z)^2\right]^{1/2}/2r^4,
  \label{eq:offB}
\end{equation}
where $a_z$ is the dipole offset, $B_d$ is the dipolar field strength\footnote{$B_d$ is defined in such a way that
the magnetic field has this strength at $z=a_z\pm1$ WD radii along the magnetic axis.}, and
\begin{equation}
  r^2 = x^2 + y^2 + (z-a_z)^2.
\end{equation}
These equations can be used to compute synthetic Zeeman line profiles given for a given $B_d$, $a_z$,
and line-of-sight inclination to the magnetic axis, $i$.
In our model we firstly generate a set of 10000 points randomly distributed over the projected surface of the WD.
At each point the magnetic field strength is evaluated using equation~\eqref{eq:offB}, accounting for
the inclination of the magnetic axis, and the corresponding Zeeman line profiles are generated
as a sum of 3 Gaussians separated according to equation~\eqref{eq:Bsep}.
These profiles are then coadded with weights proportional to the limb darkening corresponding to the location on the star.
We use limb darkening coefficients appropriate for a $6000$\,K, $\log g=8$ WD from
\citet{gianninasetal13-1}, adopting the logarithmic limb darkening law described therein.

We fitted the above model to the Zeeman lines in SDSS1536$+$4205
using the affine invariant MCMC sampler, \textsc{emcee}.
Replacing $B_S$ with $B_d$ and including the inclination and dipole offset, increased the number of free parameters in the fit
(compared with \S\ref{average} to 9.
Uninformed priors were used for variables with a physically constrained range.
E.g. $-1<a_z<+1$ forces solutions with the dipole centre confined within the stellar surface.
For the line-of-sight inclination to the magnetic axis, the prior distribution $P(i) \propto \sin i$ was employed.

\begin{figure}
  \centering
  \includegraphics[angle=90,width=\columnwidth]{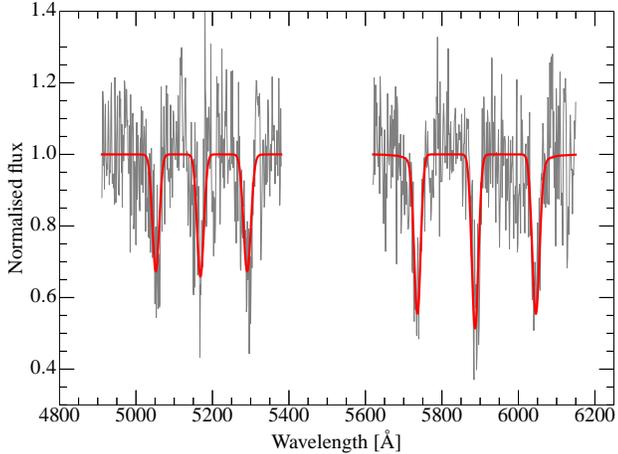}
  \caption{\label{1536_fit_2}
  Similar to Figure~\ref{1536_fit}, but fitted with an offset dipole model.
  The adopted values of $B_d$, $i$ and $a_z$ are given in Table~\ref{tab:mcmc},
  i.e. the two triplets are shown with their individual fits.
  The $\sigma$ components are now slightly less deep than the respective $\pi$ components,
  as seen in the data.}
\end{figure}

Since SDSS1536$+$4205 shows distinctly split lines of both Mg and Na, both were fitted independently.
The values for $B_d$, $i$, and $a_z$ are shown in Table~\ref{tab:mcmc},
with the resulting best fits to the spectra shown in Figure~\ref{1536_fit_2}.
While the line profiles appear similar to the simple model shown in Figure~\ref{1536_fit}, it should
be recalled that we have now fitted a physical model capable of reproducing the observed narrow Zeeman lines
rather than the assumption of unbroadened lines used in \S\ref{average}.

The fit to the Mg triplet has a slightly worse reduced $\chi^2$ than to the Na triplet.
This is the result of the Mg\,\textsc{i} line having an intrinsically asymmetric profile
due to quasi-static broadening of this transition
\citep{wehrseetal80-1,koesteretal11-1}.
Therefore assuming a Gaussian profile limits the quality of the fit.
Nevertheless, the resulting parameters from the Mg and Na fits are in agreement
within their (similarly large) uncertainties.

While the inclination uncertainties permit a wide range of values within the allowed parameter space ($0$--$90$\,$^\circ$),
the results for $B_d$ and $a_z$ strongly suggest an offset dipole.
However, it should be noted that $B_d$ and $a_z$ are highly anti-correlated,
with a correlation coefficient of $-0.99$ for both fits.
The values of $a_z$ we find are well within the range of those found for SDSS DA WDs \citep{kulebietal09-1}.

\begin{table}
  \centering
  \caption{\label{tab:mcmc}Results from our MCMC fits to SDSS1536$+$4205 assuming an offset dipole field structure.
  Quoted values and uncertainties errors are, respectively, the 50th, 15.9/84.1th percentiles of the posterior
  probability distributions. Reduced $\chi^2$ values are calculated using the median for each parameter.}
  \begin{tabular}{lcc}
    \hline
    \hline
    Parameter                 & Mg                    & Na                   \\
    \hline
    \vspace{2mm}
    $B_d$ (MG)                & $18.6_{-1.2}^{+2.2}$  & $20.0_{-1.5}^{+1.6}$ \\
    \vspace{2mm}
    inclination ($^{\circ}$)  & $31_{-15}^{+14}$      & $41_{-18}^{+16}$     \\
    \vspace{2mm}
    $a_z$ $(\%R_\mathrm{WD})$ & $-23.9_{-7.1}^{+4.3}$ & $-28.8_{-4.3}^{+5.5}$\\
    reduced $\chi^2$          & 1.25                  & 0.97                 \\
    \hline
  \end{tabular}
\end{table}

The fit values of $B_d$, $i$, and $a_z$ have the following physical interpretation:
An offset dipole leads to a strong field emerging at one of the poles of SDSS$1536$+$4205$,
($\approx 50$\,MG according to equation~\eqref{eq:offB}), with the opposite hemisphere exhibiting
a very uniform field strength of $9.6$\,MG as in Table~\ref{tab:mag_DZ}.
The value of inclination and sign of $a_z$ imply that most of the WD surface with high fields is obscured when viewed from the Earth,
with only a small amount of this region entering the limb of the star.
Hence the $\sigma$ components of the Zeeman lines are broadened only slightly with their depths reduced by a few percent.

\section{Notes on individual objects}
\noindent\textbf{SDSS0157$+$0033}. The value of $B_S$ given by \citet{schmidtetal03-1} is 3.7\,MG,
whereas we obtain a value of $3.49\pm0.05$\,MG using the same SDSS spectrum.
\citet{schmidtetal03-1} arrive at this value from measuring both the Mg and Na lines.
For the Na triplet, only two of the Zeeman components can be identified in the noisy spectrum,
and their centres are difficult to locate.
For the Mg triplet, where all three Zeeman components are well resolved, we have noticed that \citet{schmidtetal03-1}
report a position for the $\sigma^-$ line (5128\,\AA) which is about 7\,\AA\ bluer than we measure leading to an overestimate of $B_S$
We therefore suggest $3.49\pm0.05$\,MG as a revised value for this star's surface averaged magnetic field strength,
which we quote in Table~\ref{tab:mag_DZ}.

\smallskip\noindent\textbf{SDSS0735$+$2057} exhibits an extremely broad Mg feature,
and close examination reveals Zeeman split lines at its base.
None of the other 78 DZ in our sample shows an Mg line like this.
It seems likely that the unusual Mg profile is a direct result of the magnetic field partially splitting an already broad feature.
SDSS2325$+$0448 has a similar surface-averaged field strength and so one might suspect a similarly broadened Mg triplet,
which is not observed.
The stark difference between their line profiles can be explained by higher metal abundances and lower \Teff\ for SDSS0735$+$2057.
This increased opacity that the magnetic induces in the Mg feature will need to be considered
when calculating chemical abundances (Hollands et al. in prep).
The formal errors on $B_S$ (Table~\ref{tab:mag_DZ}) are surprisingly small,
at the level of ${\sim}1$\,percent, however the values obtained from the Mg and Na lines for this star agree to within
1.3$\sigma$ ($6.24\pm0.11$\,MG and $6.08\pm0.07$\,MG respectively).
We adopt the weighted mean of these as the measured value in Table~\ref{tab:mag_DZ}.

\smallskip\noindent\textbf{SDSS1536$+$4205} has the strongest magnetic field of any known DZ,
with $B_S=9.59\pm0.04$\,MG.
We also obtain very consistent field measurements between the Mg and Na lines,
$9.57\pm0.08$\,MG and $9.60\pm0.04$\,MG respectively, indicating the uncertainties are not underestimated.
Again the final value presented in Table~\ref{tab:mag_DZ} is the weighted average of these two independent measurements.
The Zeeman split lines of Mg and Na for this WD are well resolved from the noise,
yet do not show any significant magnetic broadening, 
as discussed in detail in \S\ref{offset_dipole} and \S\ref{topology}.
We also note that this object may show an emission line at approximately 8400\,\AA.
This peculiar feature does not coincide with any sky line, and peaks at $4\sigma$ above the continuum.
Additionally, the feature is visible in multiple SDSS sub-spectra.
Checking spectra from adjacent fibers observed on the same night and plate does not indicate flux contamination from other
fibers, and neither does inspection of the SDSS images reveal any nearby bright stars.
If this emission is real, the most plausible identification is O\,\textsc{i}.
We note that a few cool DQ WDs show oxygen emission lines \citep{provencaletal05-1},
the origin of which is still uncertain.
A higher S/N spectrum will be required to determine whether this feature in SDSS is genuine.

\section{Discussion}
\subsection{Magnetic incidence}
We have amassed a sample of 79 DZ WDs with colours that place them below the main sequence in the \ugr\ plane.
We identify 10 of these to be magnetic (Table~\ref{tab:mag_DZ}),
leading to an observed incidence of $13\pm4$\,percent purely for this sample.
Our selection procedure only uses colour, redshift and proper-motions.
Since colours will not be significantly altered by the presence of a magnetic field\footnote{
The dominant effects on the location in ($u-g$, $g-r$) colour space are the effective temperature and
the atmospheric abundance of Ca which causes strong opacity in the $u$-band.},
we believe this selection procedure to be unbiased towards (or against) selection of magnetic objects.
This is not to say this sample is free from biases.
On the contrary, various selection effects, which we discuss in the following sections,
suggest that our measured magnetic incidence is only a lower limit.

We report DZ to be magnetic only where we confidently detect Zeeman splitting.
The major limitation in detecting splitting of spectral lines are S/N and spectral resolution.

\begin{figure}
  \centering
  \includegraphics[angle=90,width=\columnwidth]{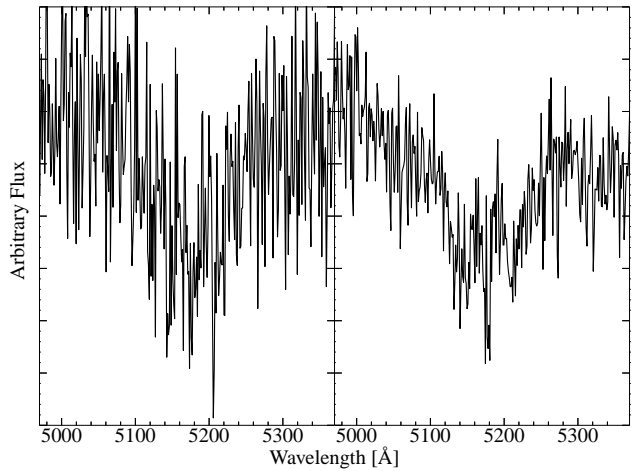}
  \caption{\label{1152Mg}
  Mg triplets for both SDSS spectra of SDSS1152$+$1605.
  The Zeeman splitting on the earlier spectrum (left) is very ambiguous,
  yet well resolved in the newer BOSS spectrum (right).}
\end{figure}

The vast majority of our 79 DZ have S/N ratios of 5--6.
While this is sufficient to identify the pressure broadened absorption features characteristic of cool DZ,
detecting fields below $2$\,MG is not possible.
For example, SDSS1152$+$1605 was identified from a spectrum observed with the SDSS spectrograph \citep{koesteretal11-1}
at a S/N ratio of ${\sim}5$.
Although \citet{koesteretal11-1} speculated that this object may be magnetic, the quality of the data available
to them was insufficient for a firm conlusion.
Its newer BOSS spectrum clearly reveals Zeeman splitting (S/N ratio of 9).
We show a comparison of these two spectra in Figure~\ref{1152Mg}.

\begin{figure}
  \centering
  \includegraphics[angle=90,width=\columnwidth]{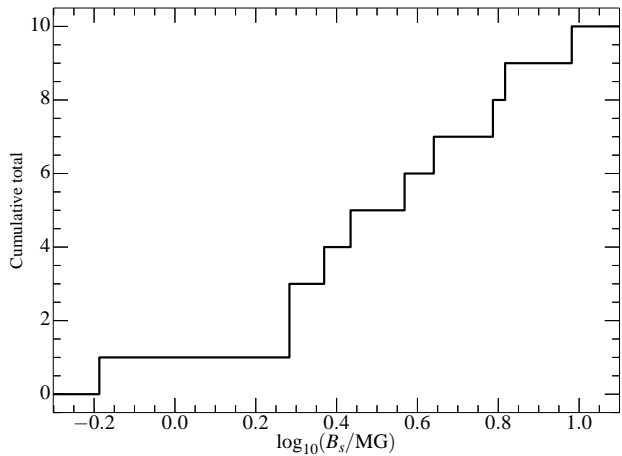}
  \caption{\label{cdf}
  Cumulative total of magnetic objects from our sample versus the log of magnetic field strength.
  The increase is seen to be approximately linear in $\log(B_S)$ above 2\,MG.}
\end{figure}

Inspecting the cumulative distribution in field strengths (Figure~\ref{cdf}) demonstrates the difficulty
in detecting magnetism for $B_S < 2$\,MG. 
Above ${\sim}1.9$\,MG, the distribution is approximately linear in $\log(B_S)$,
which is consistent with the distribution seen in other WD samples \citep{kawkaetal07-1}.
Below this value the only MWD found is SDSS1330$+$3029 ($B_S=0.65$\,MG),
which is made possible by the exceptional S/N of its spectrum.
This suggests that several objects with lower quality spectra may have magnetic fields between
$0.65$ and $1.9$\,MG.

As for resolution, the BOSS instrument has a resolving power of approximately $2000$ \citep{smee2013}.
This implies a minimum detectable field via Zeeman splitting of a few hundred kG,
and so SDSS1330$+$3029 is representative of the lowest detectable field
for DZ in SDSS.

\subsection{Magnetic field origin and evolution}
\label{origin}
While the true incidence of magnetism among WDs on the whole is still widely debated, estimates between 5--10\,percent are common
for isolated degenerate objects \citep{wickramasinghe+ferrario00-1,liebertetal03-1,sionetal14-1}.
It has been suggested that older (cooler) WDs exhibit a higher incidence of magnetism \citep{kawka+vennes14-1,liebertetal03-1},
which at face value is supported by the large fraction of cool MWDs in our sample.
However, we can at present not exclude that the high incidence of magnetism is linked
to the presence of metals in the atmospheres of the cool DZ, e.g. through (merger) interaction with planets.

In addition, the origin of magnetic fields among WDs remains under discussion with the most plausible mechanisms proposed being:

\begin{enumerate}
\item From the intial-to-final mass relation \citep{catalanetal08-2}
and main sequence lifetime as a function of stellar mass,
most of the known WDs are thought to have evolved from A and B type stars.
These stars are known to exhibit magnetic fields \citep{angel81-1,wickramasinghe+ferrario00-1,neineretal14-1},
with the peculiar Ap and Bp stars having comparatively higher fields.
As the star evolves off of the main sequence it is expected that the magnetic flux of the progenitor star is conserved
and so the change in stellar radius amplifies the surface field, i.e. $B_\mathrm{WD}/B_\mathrm{MS} = (R_\mathrm{MS}/R_\mathrm{WD})^2$.
This is known as the fossil field hypothesis,
and can produce WDs with field strengths in the observed range
\citep{woltjer64-1,angeletal70-1,angel81-1,wickramasinghe+ferrario00-1}.

Ohmic decay is expected to cause magnetic fields to decrease in strength with time.
However, the timescale for this is expected to be of the order $10^{10}$\,yr
due to the high electrical conductivity in the degenerate cores of WDs \citep{wendelletal87-1}.
Therefore the fossil field hypothesis is not unreasonable for describing the field origin in
the old WDs we identify in this work.

However, recent estimates of magnetic incidence exceeding 10\,percent \citep{liebertetal03-1,kawkaetal07-1,sionetal14-1},
challenge the fossil field hypothesis.
The space density of Ap/Bp stars is insufficient to account for all the known MWDs with $B_S \ga 1$\,MG \citep{kawka+vennes04-1},
and so at least one other evolution channel is required for producing MWDs.

\item \citet{toutetal08-1} suggested that WDs with $B_S>1$\,MG are the products of an initial binary origin.
Stellar evolution of one of the binary components can lead to a common envelope (CE) stage.
It is during this phase that a magnetic dynamo may be generated within the CE.
The resulting field then persists beyond the lifetime of the CE, within the now close binary or merged single object.
For a close binary, a merger may take place later.

The binary origin of these highly magnetised WDs naturally leads to higher masses than the canonical 0.6\,\Msun\
for non-magnetic WDs,
compatible with the observation that MWDs are typically more
massive than non-magnetics \citep{liebertetal88-1,liebertetal03-1}.

However, a binary origin would in our case raise questions about how these WDs come to be polluted by material from a remnant
planetary system.
This model need not be constrained only to stellar binaries.
\citet{nordhausetal11-1} suggested that the engulfment of gaseous planets or brown dwarf companions
during the asymptotic giant branch (AGB) phase could also lead to magnetic dynamo generation and eventually
a high field mwd.
in this scenario, mwds would not be expected to have higher masses than non-magnetics,
but it would allow for evolved planetary systems which later pollute the wd with metals.
\citet{farihietal11-2} identified a cool (\Teff$=5310$\,k) magnetic
($B\simeq120$\,kG) DAZ white dwarf,
and speculate on that basis that the WD underwent a CE with a closely orbiting gas giant planet during the
progenitor star's AGB phase, leading to the emergence of a magnetic field.
If this is indeed the mechanism from which magnetic fields are produced in DZ, it may explain
the particularly high magnetic incidence found in our sample.

Unlike the fossil field hypothesis, the giant planet CE scenario would be correlated
with the presence of metals in the atmospheres of WDs, where the metal lines are an indicator of an evolved planetary system.
Therefore, if DC white dwarfs, which originate from the same stellar population as DZ \citep{farihietal10-2},
have a significantly different distribution of magnetic fields, then this would present a compelling case for
the CE hypothesis.

\item An alternative origin for magnetism among WDs is $\alpha\omega$ dynamo generation.
For a differentially rotating WD with a convective envelope,
a magnetic dynamo may be generated at the base of the convection zone \citep{markieletal94-1}.
However this would be unlikely to produce fields on the order of $1$\,MG \citep{thomasetal95-1}, and would
lead to magnetic fields strongly aligned with the WD rotation axis which is in general not observed
\citep{latteretal87-1,burleighetal99-1,euchneretal05-1}.

\end{enumerate}

\subsection{The apparent lack of magnetism in warm DZ}
\label{warmDZ}
The largest sample of WDs identified in SDSS was presented by \citet{kleinmanetal13-1}, using SDSS DR7 spectroscopy.
In total they identified 257 DZ,
most of which are hotter than the sample we present here (\Teff$>8000$\,K),
in which case Ca H/K are usually the only metal lines detected.
Unlike in cool DZ where the broad wings of the H/K lines absorb most of the flux below 4000\,\AA\ (Figure~\ref{all_spec}),
using these lines to detect $\ga 1$\,MG fields should in principle be a trivial task.
Additionally, because of the larger sample size, there is also an abundance of these spectra where the H/K lines have good S/N ratios.
We find 64 spectra with $\mathrm{S/N} > 10$ (25\,percent), and 27 with $\mathrm{S/N} > 15$ (10\,percent).

Inspecting the Ca H/K lines of all 257 DZ did not reveal magnetic splitting for a single object.
This is in stark contrast to our fraction of 13\,percent.
One object, SDSSJ080131.15$+$532900.8, has what appear to be broadened Ca H/K lines which could indicate a magnetic field.
However the SDSS images reveal this WD to be situated ${\sim}7$\,arcsec away from a bright ($r=13.6$) M star,
which likely caused flux contamination in the DZ spectrum (obtained through a 3\,arcsec fibre).

The lack of any magnetic DZ in this sample either suggests a different set of selection effects at work
in the \citet{kleinmanetal13-1} and our own,
or that the incidence of magnetic fields in DZ differs above and below 9000\,K.
The selection procedure used by \citet{kleinmanetal13-1} was distinctly different to the methodology we employed (\S\ref{ident}).
They fitted DZ templates to a sample of WD candidate spectra,
and so magnetic objects will have evaded detection if Zeeman split lines strongly affected the $\chi^2$ of their fits.

We performed an independent check by inspecting the warm DZ identified from SDSS DR10 by \citet{gentileetal14-1}.
They selected candidate WDs from a colour-cut in the \ugr\ plane situated above the main sequence
compared to our cut in Figure~\ref{mag_ccd}, and making use of proper-motions.
All spectroscopic objects with $g<19$ bounded by this cut were visually inspected and classified into the various WD subclasses.
We inspected the H/K lines of all objects classed as DZ for splitting.
Again, of the 118 unique objects, we did not find a single star that can be convincingly claimed to be magnetic.
Since the \citet{gentileetal14-1} sample were only selected by colour and proper-motion, they are not 
biased against finding magnetic DZ as \citet{kleinmanetal13-1} might be.

\citet{kepleretal15-1} have recently published a list of new WDs from SDSS DR10 spectra including
397 objects classified as DZ (where most have \Teff$>10000$\,K).
Inspecting these 397 spectra reveals \citet{kepleretal15-1} have independently discovered 4 of the WDs that we
have shown are magnetic
(SDSS0735$+$2057, SDSS0832$+$4109, SDSS1003$-$0031, and SDSS1536$+$4205),
however, beyond these, we found no further magnetic objects.

The dearth of MWDs with \Teff$>8000$\,K in the above 3 samples may suggest that magnetic fields are preferentially
generated several Gyr after leaving the main-sequence, at least for WDs with evolved planetary systems.
If this is the case, then the origin of these fields remains a difficult question to answer.
We speculate that the gas giant/CE scenario (\S\ref{origin}) previously proposed to occur on the AGB
\citep{nordhausetal11-1,farihietal11-2},
may still remain a viable possibility at long WD cooling times.
\citet{verasetal13-1} and \citet{mustilletal14-1} have performed simulations indicating that planet-planet scattering
can still occur many Gyr after stellar evolution to the WD stage,
and show that these scattered planets will in some cases collide with central star.
We suggest such a collision with a scattered gas giant might lead to magnetic field generation.
However this scenario has two problems: 
the small fraction of systems that \citet{verasetal13-1} and \citet{mustilletal14-1} expect this to occur
for in comparison to our magnetic incidence lower limit of $13\pm4$\,percent,
and the lack of hydrogen lines observed in of the spectra in Figure~\ref{all_spec}.

\subsection{Comparison with magnetic DAZ}

For cool DZ we arrive at a magnetic incidence of $13\pm4$\,percent.
Yet if we compare this against DAZ WDs, the result is very different.
In fact very few magnetic DAZ are known at all, and their magnetic fields are not nearly as strong
as found among the DZ in this study.

\citet{kawkaetal07-1} list all magnetic WDs known up to June 2006.
Among these are the 3 previously known magnetic DZ (SDSS0157$+$0033, SDSS1214$-$0234, SDSS1330$+$3029).
However, not a single magnetic DAZ was known at that time.

Since then, 4 magnetic DAZ \emph{have} been identified, all with \Teff$<8000$\,K
\citep{farihietal11-2,zuckermanetal11-1,kawka+vennes11-1,kawka+vennes14-1},
and with the most magnetic (NLTT 10480) possessing a field of only 0.5\,MG \citep{kawka+vennes11-1}.
As with DZ (\S\ref{warmDZ}) all known magnetic DAZ have \Teff$ > 8000$\,K,
again suggesting field generation late on the WD cooling track.

Additionally, the fact that cool DAZ are not found with the same regime of magnetic field strengths as DZ
is somewhat surprising as they will have similar cooling ages,
and so the magnetic field distribution would be expected to be the same,
assuming similar progenitors.

One possible explanation is that because metal lines in DAZ
appear weaker for a given metal abundance compared with DZ,
magnetic spliting of these lines will smear them out in the continuum.
Therefore strongly magnetic DAZ would instead be classified as magnetic DA 
(where the magnetic field can still be inferred from the Balmer series).
Higher S/N spectra may reveal known magnetic DA WDs to also be metal polluted.

It is also worth noting that, to date, there are no known magnetic DBZ.
This is rather peculiar considering that metals produce stronger lines in atmospheres dominated by
helium than those dominated by hydrogen, which should easily allow the detection of fields
up to a few MG.

\subsection{Field topology}
\label{topology}
Our results from \S\ref{offset_dipole} showed that an offset dipole topology
provides a reasonable explanation for the minimal magnetic broadening seen in the Zeeman lines of some of the WDs
shown in Figure~\ref{all_spec}.

In this scenario, SDSS1536$+$4205 has a dipole offset away from the Earth leading to the distribution of
observed field strengths appearing sharply peaked at $9.59$\,MG.
It follows from this model that the opposite, invisible hemisphere of the star exhibits a large gradient in field strengths with
a strongly magnetic spot (${\sim}50$\,MG) emerging at the pole.
If the sign of $a_z$ was reversed,
the strong gradient in the field across the visible hemisphere would have major observational consequences.
The $\sigma$ components in the Mg and Na triplets would be magnetically broadened
to the extent of reducing their depth to only ${\sim}15$\,percent of the $\pi$ component.
Therefore, identifying the magnetic nature of the star would require a S/N ratio of at least 40 (for a $3\sigma$ detection).

If the offset dipole model is the correct interpretation for the narrow Zeeman lines,
this has a profound consequence for the incidence of magnetism in cool WDs.
As discussed above, of these 10 MWDs in our sample, SDSS0157$+$0033, SDSS1003$-$0031, SDSS1536$+$4205, and SDSS2325$+$0448
all have Zeeman triplets that could arguably be explained by the offset dipole model.
If cool MWDs have a tendency for their dipoles to be offset, and if all 4 of the above stars are viewed 
with their dipoles offset \emph{away} from us,
then statistically this implies that several of the other WDs within our full sample should have dipoles offset \emph{towards} the Earth.
The $\sigma$ components of their Zeeman lines would be broadened to the point that they cannot be distinguished
at the low S/N of the SDSS spectra, and so they would not been identified as magnetic.
Thus, if true, the offset dipole scenario increases the selection bias against identifying magnetism in cool WDs.

An alternative explanation for the narrowness of the Zeeman lines could also come from a non-uniform
distribution of metals across the surface of the WD.
If for instance the accreted material accumulated at the poles, the resulting spectrum would exhibit splitting
consistent with only the polar field strength.
However, to reproduce the minimal magnetic broadening we observe, the metals would have to be constrained to
such a small region that the resulting Zeeman triplets would have negligible depths.
Alternatively, metals confined to the magnetic equator would also produce a spectrum showing a small range of field
strengths, but would be able cover a much greater portion of the visible surface without significant magnetic
broadening. However, equatorial accretion would necessitate at least a quadrapolar field.

\citet{metzgeretal12-1} considered accretion of metals onto a MWD.
If the sublimation radius of the WD is smaller than the Alfv\'en radius (true for our MWD sample),
then material is expected to accrete along the magnetic field lines as soon as it enters an ionised state.
However, \citet{metzgeretal12-1} also discuss a potential caveat to this scenario.
The presence of dust grains mixed within the gaseous disc may inhibit the ionisation of the gas component,
and so even a strong magnetic field may have little influence over the accretion flow of rocky debris.

It is possible to distinguish between the two scenarios of non-uniform metal accretion and an offset dipole
by considering the rotation of magnetic WDs.
\citet{brinkworthetal13-1} showed that isolated magnetic WDs have typical rotation periods of hours to days.
Many of the WDs they observed in their sample have effective temperatures that fall into the range we study here.
In general the magnetic axis of MWDs are not aligned with the rotation axis
\citep{latteretal87-1,burleighetal99-1,euchneretal05-1},
therefore by taking spectra at multiple epochs, one would expect to see variation in the Zeeman line profiles
of these stars.

For non-uniformly distributed metals, the projected area of the metal-polluted region would phase with the rotation period.
Spectroscopically, this would be seen as a reduction in depth of all Zeeman lines in the triplet proportional
to the change in projected area.

For an offset dipole, the effect of rotation would be to bring more/less of the concealed, highly-magnetised pole into view,
leading to increased/decreased broadening of the $\sigma$ components of the Zeeman triplet.
However the $\pi$ component would remain unchanged in depth.

\section{Summary}
We have identified a sample of 79 DZ with \Teff$<9000$\,K of which \emph{at least} 10 possess magnetic fields in the range 0.5--10\,MG.
This implies a minimum incidence of $13\pm 4$\,percent which is substantially higher than for young hot DAs.
Accounting for various sources of bias, such as poor signal-to-noise spectra
and that these objects are identified from a magnitude limited sample, suggests that
the true incidence of magnetism in cool DZ is almost certainly higher.
We also note the narrowness of the Zeeman lines in several of these objects and show
that these are most likely the result of a complex field topology such as an offset dipole.
The simultaneous release of SDSS DR11 and DR12 will provide a plethora of new DZ spectra from which
improved statistics can be calculated.

\section*{Acknowledgements}
M.H. gratefully acknowledges N. Gentile Fusillo for use of his SDSS spectroscopic catalogue,
and T. Marsh for informative discussions on MCMC fitting.

The research leading to these results has received funding from the European Research Council
under the European Union’s Seventh Framework Programme (FP/2007-2013) / ERC Grant Agreement n. 320964 (WDTracer).

This research has also received funding from the University of Warwick Chancellor's Scholarship.

This work makes use of data from SDSS.
Funding for SDSS-III has been provided by the Alfred P. Sloan Foundation,
the Participating Institutions, the National Science Foundation,
and the U.S. Department of Energy Office of Science.
The SDSS-III web site is http://www.sdss3.org/.

\bibliographystyle{mn_new}
\bibliography{aamnem99,aabib}

\bsp
\label{lastpage}
\end{document}